\documentclass[aip,pop,numerical,preprint,11pt]{revtex4-1}	

\usepackage{amsbsy}
\usepackage{amsmath}
\usepackage{graphicx}

\newcommand{\bref}[1]{(\ref{#1})}											
\newcommand{\dd}{\text{d}}														
\newcommand{\del}{\partial}														
\newcommand{\Deltapol}{\Delta_{\text{pol}}}						
\newcommand{\Deltaprime}{\Delta^{\prime}}							
\newcommand{\dq}[1]{\textquotedblleft{#1}\textquotedblright}	
\newcommand{\EcrossB}{\vc{E}\times\vc{B}}							
\newcommand{\eref}[1]{Eq. (\ref{#1})}									
\newcommand{\fref}[1]{Fig. \ref{#1}}									
\newcommand{\FSave}[1]{\left\langle{#1}\right\rangle_{\Omega}}	
\newcommand{\grad}{\boldsymbol{\nabla}}								
\newcommand{\lambdaB}{\lambda B}											
\newcommand{\Maxe}{F_{\text{M}e}}											
\newcommand{\Maxi}{F_{\text{M}i}}											
\newcommand{\Maxj}{F_{\text{M}j}}											
\newcommand{\nuii}{{\nu}_{ii}}																
\newcommand{\pdf}[2]{\frac{\del#1}{\del#2}}			
\newcommand{\pdfat}[3]{\left.\frac{\del#1}{\del#2}\right|_{#3}}
\newcommand{\pl}{_{\|}}																
\newcommand{\rref}[1]{Ref. \onlinecite{#1}}					
\newcommand{\tdf}[2]{\frac{d#1}{\dd#2}}								
\newcommand{\thetaave}[1]{\left\langle{#1}\right\rangle_{\theta}}
\newcommand{\uprime}{^{\prime}}												
\newcommand{\vc}[1]{\mathbf{#1}}											

\begin{document}

\title{Collision frequency dependence of polarization current in neoclassical tearing modes}

\author{K. Imada}
\author{H.R. Wilson}
\affiliation{York Plasma Institute, Department of Physics, University of York, Heslington, York YO10 5DQ, U.K.}

\date{\today}

\begin{abstract}
The neoclassical polarization current, generated when a magnetic island propagates through a tokamak plasma, is believed to influence the initial stage of the neoclassical tearing mode evolution. Understanding the strength of its contribution in the relevant plasma collision frequency regimes for future tokamaks such as ITER is crucial for the successful control and/or avoidance of the neoclassical tearing mode. A nonlinear drift kinetic theory is employed to determine the full collision frequency dependence of the neoclassical polarization current in the small island limit, comparable to the trapped ion orbit width. Focusing on the region away from the island separatrix (where a layer with a complex mix of physics processes exists), we evaluate for the first time the variation of the neoclassical ion polarization current in the transition regime between the analytically tractable collisionless and collisional limits. In addition, the island propagation frequency-dependence of the neoclassical polarization current and its contribution to the island evolution is revealed. For a range of propagation frequencies, we find that the neoclassical polarization current is a maximum in the intermediate collision frequency regime analyzed here - a new and unexpected result.

\end{abstract}

\maketitle

\section{Introduction}
\label{introduction}

A tokamak plasma is subject to a number of magnetohydrodynamic (MHD) instabilities that can limit the performance of the toroidal fusion device. In tearing mode instabilities, filamentation of the plasma current density along equilibrium magnetic field lines forms a chain of magnetic islands at a rational surface, and leads to a corrugation of flux surfaces in its vicinity. Within the island, the enhanced radial transport of particles and heat reduces the radial pressure gradient. This results in a reduction of the core plasma pressure, which degrades the plasma confinement. According to the theory of single-fluid resistive MHD, the evolution of a magnetic island is characterized by the rate of change in its width, which is proportional to the parameter $\Deltaprime$ \cite{1973PoF16-1903}: a measure of the free energy stored in the plasma current density for the magnetic reconnection to occur. For $\Deltaprime >0$, the island is predicted to grow. The neoclassical theory of tearing modes incorporates the effects of toroidal geometry in the layer surrounding the rational surface. One contribution to the current that influences the island evolution is the bootstrap current, which is proportional to the radial pressure gradient. Because of the pressure gradient flattening, the bootstrap current is suppressed inside the island region. This perturbation in the bootstrap current enhances the original filamentation of the plasma current and hence drives the island growth \cite{1985UWPR85-5, 1986PoF29-899}. The neoclassical tearing modes (NTMs) are characterized by this enhanced drive for the island growth, whose strength diminishes with increasing island width, $w$. The result is that $w$ typically saturates at a substantial fraction of the tokamak minor radius, $r$. In toroidal geometry, the curvature of the magnetic field lines provides a stabilizing contribution to the island evolution equation \cite{1975PoF18-875, 2001PoP8-4267}. This curvature effect is also found to be proportional to the pressure gradient and diminishes with the island width. Thus it can be thought of as an effect that reduces the bootstrap drive (except for sufficiently small islands \cite{2001PoP8-4267}), though the effect is typically small for the large aspect ratio tokamaks we consider in this paper.

\begin{figure}
	\includegraphics{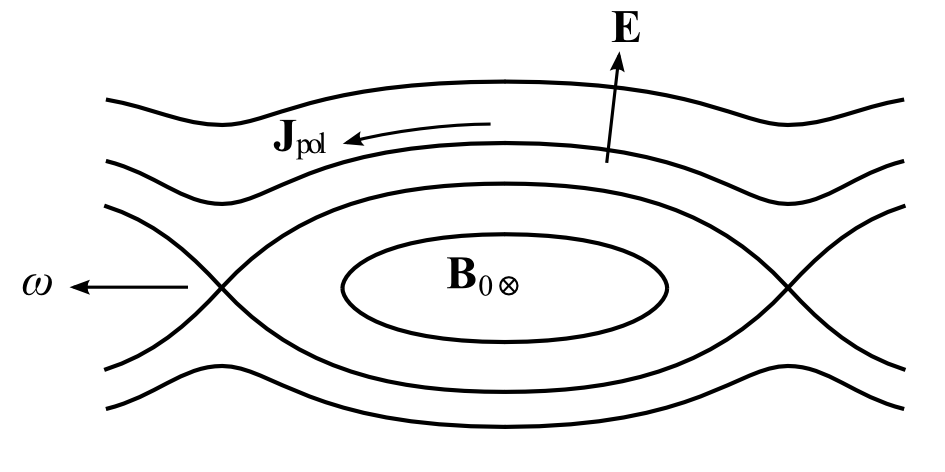}
	\caption{\label{Jpol} A cross section of a magnetic island indicating the directions of the electric field perturbation ($\vc{E}$), equilibrium magnetic field ($\vc{B}_{0}$), island propagation ($\omega$) and the polarization current ($\vc{J}_{pol}$).}
\end{figure}

There is substantial experimental evidence for the existence of a threshold mechanism \cite{1995PRL74-4663, 1998NF38-987, 2001NF41-197, 2002PPCF44-1999, 2010PPCF52-104041}, whereby a sufficiently small \dq{seed} island (typically of $O(1\text{cm})$) does not grow to a large saturated island, but rather shrinks away. However, there is no concrete theoretical framework to explain the observed level of the threshold width for the island width and provide quantitative predictions. One possible mechanism is the effect of finite cross-field transport of particles and heat in the vicinity of the island separatrix \cite{1995PoP2-825, 1997PoP4-2920}. A consequence of the finite radial transport is that the pressure gradient is not completely flattened across the island width, and then the bootstrap drive for the island width is reduced. Another possible source of the threshold mechanism is the polarization current, which is induced when the island is in relative motion with the bulk of the plasma. A number of works have considered the finite ion Larmor radius (FLR) effect on tearing mode evolution in sheared slab geometry \cite{1989SJPP15-667, 1993PPCF35-657, 2006PoP13-122507, 2006PPCF48-1647, 2001PRL87-215003, 2001PoP8-2835}, which can give rise to the polarization current when the island width is comparable to the ion Larmor radius, $\rho_{Li}$. However, whether this FLR effect can stabilize the island depends on the plasma conditions and, in particular, the island propagation frequency. In toroidal geometry, the effect of finite trapped ion orbit width, or the ion \dq{banana width}, $\rho_{bi}$, can also generate a polarization current. Because $\rho_{bi} \gg \rho_{Li}$, this neoclassical contribution dominates the polarization current except perhaps in the vicinity of the island separatrix, where the FLR effect is also likely to be important. The origin of the neoclassical polarization current can be understood as follows. As the magnetic island propagates through the plasma, ions and electrons respond differently; the ion response is dominated by the $\EcrossB$ drift, whereas the parallel transport dominates the electron response. However, because of the difference in ion and electron banana widths ($\rho_{bi}\gg \rho_{be}$), the orbit-averaged $\EcrossB$ drifts of trapped ions and electrons differ, resulting in a net current perpendicular to the magnetic field lines (see \fref{Jpol}). This is the neoclassical polarization current, which in turn generates a parallel current perturbation to ensure that $\grad.\vc{J}=0$, where $\vc{J}$ is the current density. It is this parallel current perturbation that contributes to the island evolution. Past analytical works \cite{1995PoP2-1581, 1996PoP3-248, 2009PPCF51-105010} show that this neoclassical polarization current contribution depends strongly on the plasma collision frequency regime; it is $O(\epsilon^{3/2})$ smaller in the collisionless limit ($\nuii\ll \epsilon\omega$), compared to the collisional limit ($\nuii\gg \epsilon\omega$). Here, $\nuii$ is the ion-ion collision frequency, $\epsilon=r/R$ is the inverse aspect ratio ($r$ and $R$ are the minor and major radii respectively) and $\omega$ is the island propagation frequency in the $\EcrossB$ rest frame. In this paper, we aim to develop the nonlinear drift kinetic theory of Refs. \onlinecite{1996PoP3-248} and \onlinecite{2009PPCF51-105010} further to determine the full collision frequency dependence of the neoclassical polarization current.
%

As discussed above, the neoclassical polarization current is seeded by the trapped particles (through the finite ion banana width effect). Collisions transfer the current perturbation carried by the trapped particles to the passing particles at the rate $\nuii/\epsilon$. In the high collision frequency limit $\nuii/\epsilon\gg\omega$ (but collisionality, $\nu_{*}<1$), the collisional momentum transfer takes place sufficiently quickly that the time variation of the trapped particle response is resolved, and the polarization current is communicated to the passing particles and amplified in the process. On the other hand, in the low collision frequency limit ($\nuii\ll\epsilon\omega$) the collisional transfer does not resolve the time variation and the polarization current remains at the seed level, which is $O(\epsilon^{3/2})$ smaller than that in the collisional limit.

Taking into account the contributions discussed so far, the island evolution equation can be written in the form of the \dq{modified Rutherford equation}:
%
\begin{equation}
	\label{mRE}
	\frac{\tau_{R}}{r^{2}}\tdf{w}{t} = \Deltaprime +\Delta_{bs} +\Delta_{GGJ}
		+\Delta_{pol} +\Delta_{layer}.
\end{equation}

\noindent Here, $\tau_{R}$ is the resistive time scale, $\Delta_{bs}$ denotes the bootstrap drive for the island growth and $\Delta_{GGJ}$ denotes the contribution from the curvature effect. As discussed in detail below, we separate the contribution from the neoclassical polarization current into two parts: $\Deltapol$ denotes the contribution from the neoclassical polarization current away from the island separatrix (which we shall call the \dq{external polarization current}) and $\Delta_{layer}$ denotes the contribution from the narrow layer that surrounds the island separatrix (the \dq{layer polarization current}). Our interest in this paper is focused on the term $\Deltapol$. We characterize the strength of the neoclassical polarization current by introducing the dimensionless parameter $g(\nuii,\epsilon,\omega)$ and write
%
\begin{equation}
	\label{Deltapol}
	\Delta_{pol} = -g(\nuii,\epsilon,\omega)\left(\frac{L_{q}}{L_{p}}\right)^{2}
		\left(\frac{\rho_{bi}}{w}\right)^{2}\frac{\beta_{\theta}}{w},
\end{equation}

\noindent where $L_{q}= q/(dq/dr)$, $L_{p}= p/(dp/dr)$ and $\beta_{\theta}$ is the plasma poloidal $\beta$. Analytic results from \rref{1996PoP3-248} in the collisionless limit and Refs. \onlinecite{1995PoP2-1581} and \onlinecite{2009PPCF51-105010} in the collisional limit provide
%
\begin{equation}
	\label{ganalytic}
	g(\nuii,\epsilon,\omega) = \left\{
	\begin{array}{lc}
		1.64\epsilon^{3/2}f(\nuii/\epsilon\omega)\omega(\omega-\omega_{*pi})/\omega_{*i}^{2}
			\ & \nuii/\epsilon\omega \ll 1, \\
		\omega(\omega-\omega_{*pi}-k\eta_{i}\omega_{*i})/\omega_{*i}^{2} \ & \nuii/\epsilon\omega \gg 1,
	\end{array}
	\right.
\end{equation}

\noindent where $f(\nuii/\epsilon\omega) = 1 +\gamma\sqrt{\nuii/\epsilon\omega}$ and $\omega_{*pi}=\omega_{*i} (1+\eta_{i})$. Here, $\omega_{*i}=(mT_{i}/qq_{i}n)(dn/d\chi)$ is the ion diamagnetic frequency, $k=-1.17$ in the banana regime \cite{1981NF21-1079}, $\eta_{i}=L_{n}/L_{Ti}$ is the ratio of the density to temperature gradient length scales, $L_{n}=n/(dn/dr)$ and $L_{Ti}=T_{i}/(dT_{i}/dr)$. $\gamma$ is a weak logarithmic function of $\nuii/\epsilon\omega$ and the dependence on $\gamma\sqrt{\nuii/\epsilon\omega}$ appears from the finite collisional effects in the narrow layer around the trapped/passing boundary in pitch angle space \cite{2009PPCF51-105010}. Other works have considered the effect of collision frequency on the polarization current, including: \rref{2000PPCF42-309}, which employs the linear MHD theory; and Refs. \onlinecite{2005PRL94-205001} and \onlinecite{2005PoP12-072501}, in which the collision frequency dependence is calculated by numerical simulation. In Refs. \onlinecite{2005PRL94-205001} and \onlinecite{2005PoP12-072501}, however, only the ion response is determined and quasineutrality is not imposed to derive the self-consistent electrostatic potential, $\Phi$. As has been pointed out in Refs. \onlinecite{1996PoP3-248} and \onlinecite{2009PPCF51-105010}, the polarization current is proportional to the second radial derivative of $\Phi$, $\Phi^{\prime\prime}$. Hence it is likely to be important to work with a self-consistent form of $\Phi$, which introduces an additional complication.

There is a large contribution (formally a $\delta$-function) to $\Phi^{\prime\prime}$ from the vicinity of the island separatrix, where simplified models predict a discontinuity in the electric field; this generates a skin current contribution to the ion polarization current. This provides an additional contribution to the island evolution equation denoted by $\Delta_{layer}$ in \eref{mRE}, which is expected to be comparable in magnitude and of opposite sign to the contribution from outside the separatrix layer \cite{1997PRL78-1703, 2001PRL87-215003}, $\Deltapol$. However, the contribution from this separatrix layer requires an extended physics model to treat it, including the FLR effect \cite{2001PRL87-215003}, non-linearities in the parallel electric field and non-perturbative cross-field diffusion \cite{2010PPCF52-075008}. Consequently its contribution to the island evolution scales differently with plasma parameters to that outside the layer. It is therefore appropriate to separate the calculation into two regions, considering the contribution to the neoclassical ion polarization current from each region separately. In this paper, we address only the contribution from outside the separatrix layer, which we called the \dq{external} polarization current [see \eref{mRE}]. This enables us to focus on the physics of the transition between the collision frequency regimes without the complicated physics of the separatrix layer. We leave the more complicated layer polarization current calculation for future research.

To summarize, the objective of this paper is to extend the drift kinetic theory developed in Refs. \onlinecite{1996PoP3-248} and \onlinecite{2009PPCF51-105010} to determine the full collision frequency dependence of the \dq{external} neoclassical ion polarization current (i.e. that away from the separatrix layer) and calculate its contribution to the island evolution. In particular, we consider the dependence of $g(\nuii,\epsilon,\omega)$ on the collision frequency in the previously unexplored intermediate regime, where $\nuii/\epsilon\omega \sim 1$. In addition, we consider the influence of the island propagation frequency $\omega$ on the contribution of the neoclassical polarization current to the island evolution. We find that whether $\Deltapol$ provides a stabilizing contribution (in this paper, this corresponds to $g>0$) depends crucially on the relative size of $\omega$ compared to the ion diamagnetic frequency, $\omega_{*i}$.

The paper is organized as follows: in Section \ref{theory}, we summarize the calculations of particle responses to the perturbed magnetic geometry, which is discussed in full in \rref{1996PoP3-248}. At leading order in $\rho_{bi}/w$, we introduce the free function associated with integration along magnetic field lines, $\bar{h}_{i}$, which carries the leading order collision frequency dependence due to the finite ion banana width effect. Constraint equations at higher order in $\rho_{bi}/w$ determine its full form. In Section \ref{calculations} we discuss the method of calculating the collision frequency dependence $g(\nuii,\epsilon,\omega)$ from the ion response to the magnetic island. Our new results for $g(\nuii,\epsilon,\omega)$ across the collision frequency range are presented in Section \ref{results}, and conclusions are drawn in Section \ref{conclusion}.

\section{Magnetic Geometry and the Drift Kinetic Equation}
\label{theory}

A tokamak plasma with a large aspect ratio ($\epsilon \ll1$) and a circular cross section is considered. We introduce a single helicity magnetic perturbation of the following form, assuming the constant-$\psi$ approximation:
%
\begin{equation}
	\label{psi}
	\psi(\xi) = \tilde{\psi}\cos{\xi},
\end{equation}

\noindent where $\xi$ is the helical angle:
%
\begin{equation}
	\label{xi}
	\xi = m\left(\theta-\frac{\phi}{q_{s}}\right).
\end{equation}

\noindent Here, $q_{s}=m/n$ is the value of the safety factor at the rational surface, and $m$ and $n$ are the poloidal and toroidal mode numbers respectively. Then, in the toroidal coordinate system $(\chi,\theta,\phi)$, where $\chi$ is the poloidal flux, $\theta$ is the poloidal angle and $\phi$ is the toroidal angle, the total magnetic field is given by:
%
\begin{equation}
	\label{B}
	\vc{B}=I(\chi)\grad\phi + \grad\phi\times\grad(\chi+\psi),
\end{equation}

\noindent where $I(\chi)=RB_{\phi}$ and $B_{\phi}$ is the toroidal component of $\vc{B}$. It is convenient to define a perturbed flux function satisfying $\vc{B}.\grad\Omega=0$:
%
\begin{equation}
	\label{fluxfunction}
	\Omega = \frac{2(\chi-\chi_{s})^{2}}{w_{\chi}^{2}} - \cos{\xi},
\end{equation}

\noindent where $w_{\chi} = RB_{\theta}w$, $B_{\theta}$ is the poloidal field and $\chi_{s}$ is the value of $\chi$ at the rational surface.

The particle responses to the magnetic island perturbation are described by the drift kinetic equation:
%
\begin{align}
	\label{DKE}
	\pdf{f_{j}}{t} +& v\pl\nabla\pl f_{j} +\vc{v}_{E}.\grad f_{j} +\vc{v}_{b}.\grad f_{j} \nonumber\\
	&+\frac{q_{j}}{m_{j}}\frac{v\pl E\pl}{v}\pdf{f_{j}}{v}
		-\frac{q_{j}}{m_{j}}\frac{\vc{v}_{b}.\grad\Phi}{v}\pdf{f_{j}}{v} = C(f_{j}),
\end{align}

\noindent where $\vc{v}_{E}$ is the $\EcrossB$ drift, $\vc{v}_{b}=-v\pl\vc{b}\times \grad(v\pl/\omega_{cj})$ is the magnetic drift and $\omega_{cj} =q_{j}B/m_{j}$ is the gyrofrequency. $\Phi$ is the electrostatic potential perturbation to be obtained from quasineutrality, and $C$ is the model collision operator. In the coordinate system $(\chi,\theta,\xi)$, the parallel derivative operator is given by:
%
\begin{equation}
	\label{grad||}
	\nabla\pl\equiv \frac{\vc{B}.\grad}{B}
		= \frac{1}{Rq}\pdfat{}{\theta}{\xi,\chi} +k\pl\pdfat{}{\xi}{\Omega,\theta},
\end{equation}

\noindent where
%
\begin{equation}
	\label{k||}
	k\pl = m\frac{(\chi-\chi_{s})}{Rq}\frac{q_{s}\uprime}{q_{s}}
\end{equation}

\noindent and $q_{s}\uprime = (dq/d\chi)_{\chi=\chi_{s}}$. As discussed in the introduction, we restrict our analysis to the region outside the separatrix layer, which corresponds to $\Omega>1$. The parallel derivative operator can then be annihilated by introducing the following averaging operators:
%
\begin{equation}
	\label{FSave}
	\FSave{...} = \frac{\oint{...[\Omega+\cos{\theta}]^{-1/2}d\xi}}
		{\oint{[\Omega+\cos{\theta}]^{-1/2}d\xi}}
\end{equation}

\noindent and
%
\begin{align}
	\label{thetaavepassing}
	\thetaave{...} &= \frac{1}{2\pi}\oint{...d\theta}, \\
	\label{thetaavetrapped}
	\thetaave{...} &= \sum_{\sigma=\pm 1}\frac{1}{2\pi}\int_{-\theta_{b}}^{+\theta_{b}}{...d\theta},
\end{align}

\noindent where Eqs.\bref{thetaavepassing} and \bref{thetaavetrapped} are the $\theta$-averaging operators for the passing and trapped particles respectively, and $\theta_{b}$ is the bounce point for the trapped particles. In \eref{thetaavetrapped}, $\sigma=\pm 1$ is the sign of the parallel velocity: $v\pl = \sigma v \sqrt{1-\lambdaB}$.  $\nabla_{\|}$ in \eref{grad||} can be annihilated by the operator $\FSave{\thetaave{Rq...}}$.

The perturbed distribution function in \eref{DKE} is expressed in terms of the adiabatic and non-adiabatic parts:
%
\begin{equation}
	\label{fj}
	f_{j} = \left(1-\frac{q_{j}\Phi}{T_{j}}\right)\Maxj + g_{j},
\end{equation}

\noindent where $\Maxj$ is the Maxwellian distribution. The non-adiabatic term, $g_{j}$, is solved for each particle species by expanding it in terms of two small parameters\cite{1996PoP3-248}: $\Delta=w/r$ and $\delta_{j} = \rho_{bj}/w$, where $\rho_{bj}=\epsilon^{1/2}\rho_{\theta j}$ is the trapped particle banana width and $\rho_{\theta i}= m_{i}v_{thi}/q_{i}B_{\theta}$. Thus,
%
\begin{equation}
	\label{gj}
	g_{j} = \sum_{m,n}\delta_{j}^{m}\Delta^{n} g_{j}^{(m,n)},
\end{equation}

\noindent and we solve for the expansion terms $g_{j}^{(m,n)}$ by considering the relevant order contributions to the drift kinetic equation \bref{DKE}. For the purpose of this paper, we are interested in the collision frequency dependence of the ion response. The full derivations of the ion and electron responses are given in \rref{1996PoP3-248}. Here, we restrict our discussion to a description of the key parts of the calculation.

To leading order ($O(\delta_{i}^{0}\Delta^{0})$), the electron and ion distribution functions are given by
%
\begin{align}
	\bar{g}_{e}^{(0,0)} &= \frac{qq_{e}\Maxe}{mT_{e}}(\omega-\omega_{*e}^{T})[\chi -h(\Omega)],
		\label{ge00}\\
	\bar{g}_{i}^{(0,0)} &= \frac{\Maxi}{n}\tdf{n}{\chi}\frac{(\omega-\omega_{*i}^{T})}{\omega_{*i}}
		[\chi -h(\Omega)] \label{gi00},
\end{align}

\noindent where the bar above a quantity indicates that it is independent of $\theta$,
%
\begin{equation}
	\label{omega*T}
	\omega_{*i}^{T} = \omega_{*i}\left[1 +\left(\frac{v^{2}}{v_{thi}^{2}}-\frac{3}{2}\right)\eta_{i}
		\right],
\end{equation}

\noindent and the self-consistent electrostatic potential is
%
\begin{equation}
	\label{Phi}
	\Phi = \frac{\omega q}{m}[\chi -h(\Omega)].
\end{equation}

\noindent Here, $h(\Omega)$ is a free function that is related to the electron density profile in the vicinity of the island separatrix and can be determined from the consideration of the radial particle transport \cite{1996PoP3-248, 1992PoFB4-4072}. The $O(\delta_{i}^{1}\Delta^{0})$ equation provides:
%
\begin{equation}
	\label{g10}
	g_{i}^{(1,0)} = -\frac{Iv\pl}{\omega_{ci}}\frac{\Maxi}{n}\tdf{n}{\chi}
		\left[\frac{\omega}{\omega_{*i}} -\frac{(\omega-\omega_{*i}^{T})}{\omega_{*i}}
		\pdf{h}{\chi}\right] + \bar{h}_{i},
\end{equation}

\noindent where $\bar{h}_{i}(\Omega,\xi)$ is a free function that arises as a consequence of integration along unperturbed field lines. It is shown in \rref{1996PoP3-248} that the leading order contribution to the collision frequency dependence of the polarization current comes from this term. The explicit form for $\bar{h}_{i}$ is determined from a solubility constraint on the higher order equation. This constraint equation is obtained by averaging the $O(\delta_{i}\Delta)$ equation over the unperturbed field lines. The equation for the passing particles in the limit $\omega\gg k\pl v\pl$ (appropriate for thin islands) is:
%
\begin{align}
	\label{constraint_passing}
	-Rqk\pl\thetaave{\frac{Rq}{v\pl}\frac{\omega}{m\tilde{\psi}}\tdf{h}{\Omega}
		\pdfat{g_{i}^{(1,0)}}{\xi}{\Omega}} \hspace{25mm}\nonumber\\
	+\thetaave{\frac{Rq}{v\pl}C_{i}\left(g_{i}^{(1,0)}\right)} =0,
\end{align}

\noindent while for the trapped particles:
%
\begin{equation}
	\label{constraint_trapped}
	-Rqk\pl\frac{\omega}{m\tilde{\psi}}\thetaave{\frac{Rq}{|v\pl|}}\pdfat{\bar{h}_{i}}{\xi}{\Omega}
		+\thetaave{\frac{Rq}{|v\pl|}C_{i}\left(\bar{h}_{i}\right)} = 0,
\end{equation}

\noindent where $|v\pl| = v|\sqrt{1-\lambdaB}|$. In the collisionless limit $\nuii\ll \epsilon\omega$, the second term of \eref{constraint_passing} [or \eref{constraint_trapped}] is negligible and the first term, which describes the response to the $\EcrossB$ flow, determines $\bar{h}_{i}$. In the opposite limit $\nuii\gg \epsilon\omega$, the first term of \eref{constraint_passing} becomes negligible and the collisional effect alone determines $\bar{h}_{i}$. These analytic limits have been explored in Refs. \onlinecite{1996PoP3-248} and \onlinecite{2009PPCF51-105010}. In this paper, we consider the arbitrary collision frequency regime between the two analytically tractable limits and numerically solve the full constraint equations \bref{constraint_passing} and \bref{constraint_trapped}. In our calculation we use the following momentum-conserving model collision operator \cite{helander_sigmar192}:
%
\begin{align}
	\label{Cii}
	C_{ii}(g_{i}) = 2\nuii(v) \left[\frac{\sqrt{1-\lambdaB}}{B}\pdf{}{\lambda}
		\left(\lambda\sqrt{1-\lambdaB}\pdf{g_{i}}{\lambda}\right) \right. \nonumber\\
	\left.+\frac{v\pl\bar{u}_{\|i}}{v_{thi}^{2}}\Maxi\right]
\end{align}

\noindent for the ion-ion collisions. Here, $\nuii(v)$ is the $90^{\circ}$ deflection frequency \cite{1976PoF19-656}:
%
\begin{equation}
	\label{nuii(v)}
	\nuii(v) = \hat{\nu}_{ii} \frac{\phi(x)-G(x)}{x^{3}},
\end{equation}

\noindent where $x=v/v_{thi}$, $\phi(x)$ is the error function, $G(x)$ is the Chandrasekhar function
%
\begin{equation}
	\label{Chandrasekhar}
	G(x) = \frac{\phi(x)-x\phi\uprime(x)}{2x^{2}}
\end{equation}

\noindent and $\phi\uprime = d\phi/dx$. In \eref{Cii}, the parallel flow in the momentum conservation term is defined as
%
\begin{equation}
	\label{u||}
	\bar{u}_{\|i} = \frac{1}{n\{\nuii(v)\}}\int{d^{3}\vc{v}\ \nuii(v)v\pl g_{i}},
\end{equation}

\noindent where
%
\begin{equation*}
	\{\nuii(v)\} = \frac{8}{3v_{thi}^{5}\sqrt{\pi}}
		\int_{0}^{\infty}{e^{-v^{2}/v_{thi}^{2}}v^{4}\nuii(v)\ dv}.
\end{equation*}

\noindent Ion-electron collisions are small compared to the ion-ion collisions and are therefore neglected. In the next section, we discuss the method of solving Eqs.\bref{constraint_passing} and \bref{constraint_trapped}, and how the coefficient $g(\nuii,\epsilon,\omega)$ is determined from the solution for $\bar{h}_{i}$.

\section{Calculation of $g(\nuii,\epsilon,\omega)$}
\label{calculations}

As shown in \rref{1996PoP3-248}, the ion response provides the dominant contribution to the piece of the parallel current perturbation, $\bar{J}\pl$, which varies along magnetic field lines. This provides the contribution of the neoclassical ion polarization current to the island evolution. The leading order contribution to this parallel current perturbation comes from $g_{i}^{(1,0)}$ (which includes the collision frequency dependent part, $\bar{h}_{i}$):
%
\begin{align}
	\label{eqnforJ||}
	Rqk\pl\pdfat{\bar{J}\pl}{\xi}{\Omega} = I\frac{Rq}{\omega_{ci}}&q_{i}\frac{\omega}{m\tilde{\psi}}
		\nonumber\\
	&\times \int{d^{3}\vc{v}\ v\pl\pdf{}{\chi}\left(\tdf{h}{\Omega}\thetaave{Rqk\pl
		\pdfat{g_{i}^{(1,0)}}{\xi}{\Omega}}\right)}.
\end{align}

\noindent Integrating \eref{eqnforJ||} provides two parts for $\bar{J}\pl$: first is the parallel current perturbation arising from the neoclassical polarization current, which we identify as the part of $\bar{J}\pl$ that varies along the field lines and flux surface-averages to zero \cite{1996PoP3-248}. The second part, which is constant on a flux surface, is the flux surface average of the bootstrap current perturbation, which is determined from the perturbed ion and electron parallel flows. It is incorporated in the function of $\Omega$ following the integration of \eref{eqnforJ||}. In this paper we focus on the first part: the contribution from the neoclassical polarization current, which is determined from \eref{eqnforJ||} with the condition that this part of $\bar{J}\pl$ flux surface-averages to zero. Once $\bar{J}\pl$ is determined, the contribution of the neoclassical polarization current to the island evolution can be calculated via
%
\begin{equation}
	\label{Deltapolcalc}
	\Deltapol = \sum_{\pm} \int_{1}^{\infty}d\Omega
		\oint{\frac{\bar{J}\pl\cos{\xi}}{\sqrt{\Omega+\cos{\xi}}}\ d\xi},
\end{equation}

\noindent where the summation is over the region $\chi>\chi_{s}$ and $\chi<\chi_{s}$. As discussed in Section \ref{introduction}, the collision frequency dependence of the neoclassical polarization current and its contribution to the island evolution is described by the coefficient $g(\nuii,\epsilon,\omega)$.
%
\begin{figure}
	\includegraphics{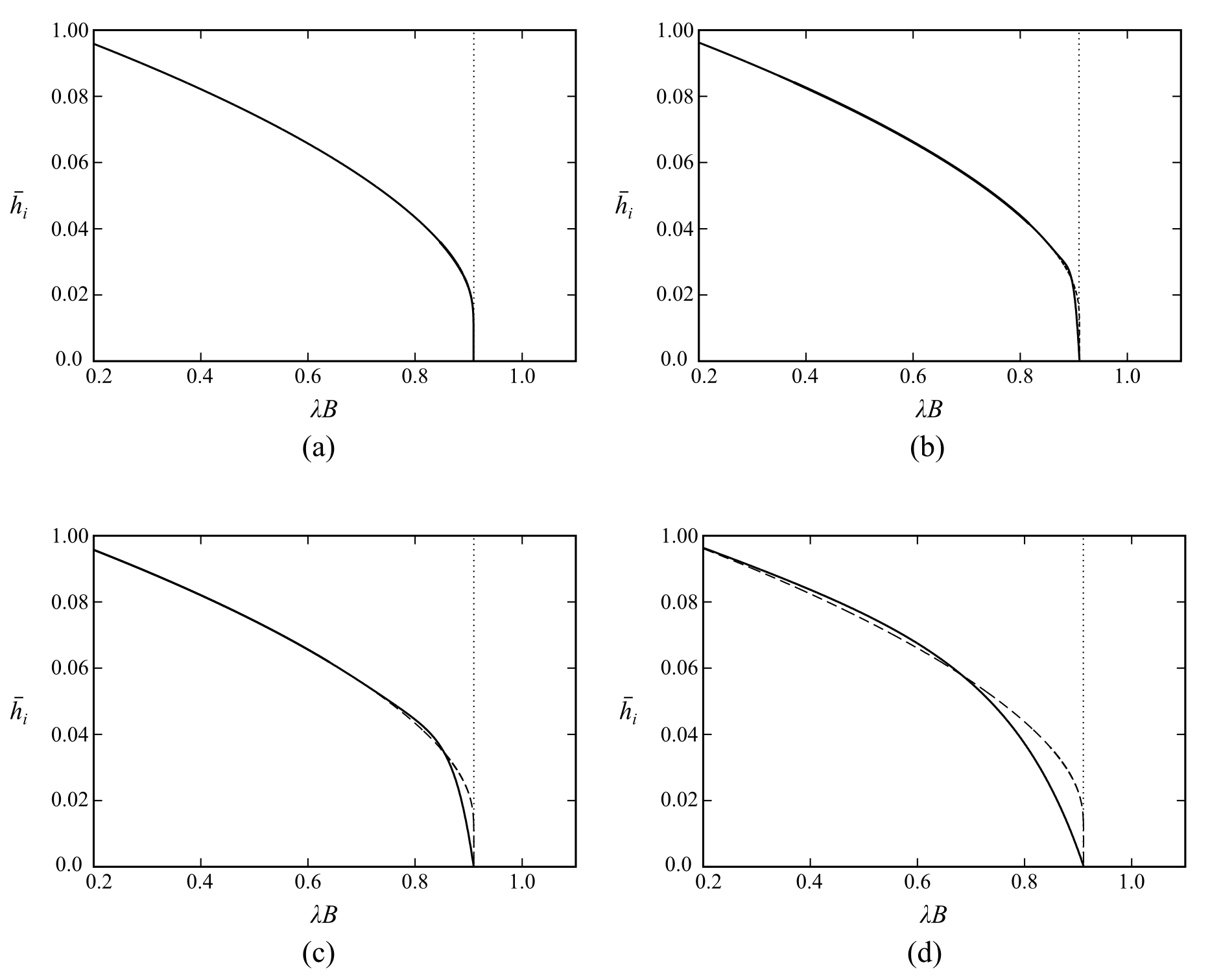}
	\caption{\label{deviation} Plots of the numerical results for $\bar{h}_{i}$ as a function $\lambdaB$ for different values of $\nuii/\epsilon\omega$, at $\epsilon=0.1$: (a) is at $\nuii/\epsilon\omega = 10^{-5}$; (b) at $\nuii/\epsilon\omega =10^{-2}$; (c) at $\nuii/\epsilon\omega =10^{-1}$ and (d) at $\nuii/\epsilon\omega =10^{0}$. The vertical dotted lines correspond to the trapped/passing boundary, $\lambda= \lambda_{c}$. The dashed curves in (b) $\sim$ (d) correspond to the collisionless result, $\nuii/\epsilon\omega =10^{-5}$, to aid comparison. Note the steep gradients that need to be resolved in the vicinity of the trapped/passing boundary at low values of $\nuii/\epsilon\omega$.}
\end{figure}

A numerical code is developed to determine $g(\nuii,\epsilon,\omega)$ from the solution for $\bar{h}_{i}$, using Eqs.\bref{g10},\bref{constraint_passing}, \bref{eqnforJ||}, \bref{Deltapolcalc} and \bref{Deltapol}. A particularly careful treatment is required for analyzing the pitch angle space; as discussed in \rref{2009PPCF51-105010}, in the total absence of collisions (i.e. when the second term of \eref{constraint_passing} becomes zero), $\bar{h}_{i}$ is discontinuous at the trapped/passing boundary. Inclusion of collisional effects smooths out this discontinuity in a narrow boundary layer (see \fref{deviation}). The pitch angle mesh needs to be closely packed to resolve this \dq{dissipation layer} that surrounds the trapped/passing boundary, in which the collisional effect is important even in the low collision frequency limit, due to steep gradients in pitch angle. A further complication is the treatment of the flow term, $\bar{u}_{\|i}$ in the momentum conserving term of the model collision operator (see \eref{Cii}). This provides an integro-differential equation, which we solve by an iterative numerical scheme. The results for $g(\nuii,\epsilon,\omega)$ are presented in the next section.

\section{Results}
\label{results}

\begin{figure}
	\includegraphics{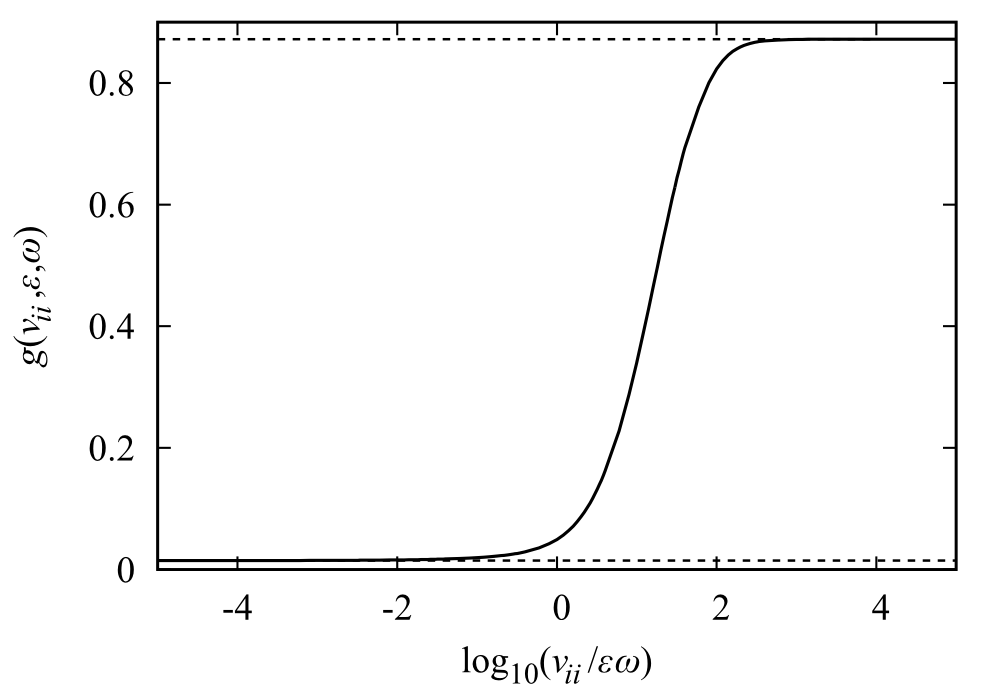}
	\caption{\label{plot1} Plot of $g(\nuii,\epsilon,\omega)$ as a function of collision frequency regime $\nuii/\epsilon\omega$ for $\epsilon=0.1$, $\eta_{i} = 1.0$ and $\omega/\omega_{*i} = 2.5$. The horizontal dotted lines represent the analytic values of $g(\nuii,\epsilon,\omega)$ in the two collision frequency limits. The plot shows that $g$ starts deviating from its collisionless value from as early as $\nuii/\epsilon\omega \sim0.1$, and this deviation is clear by $\nuii/\epsilon\omega=1$.}
\end{figure}

\begin{figure}
	\includegraphics{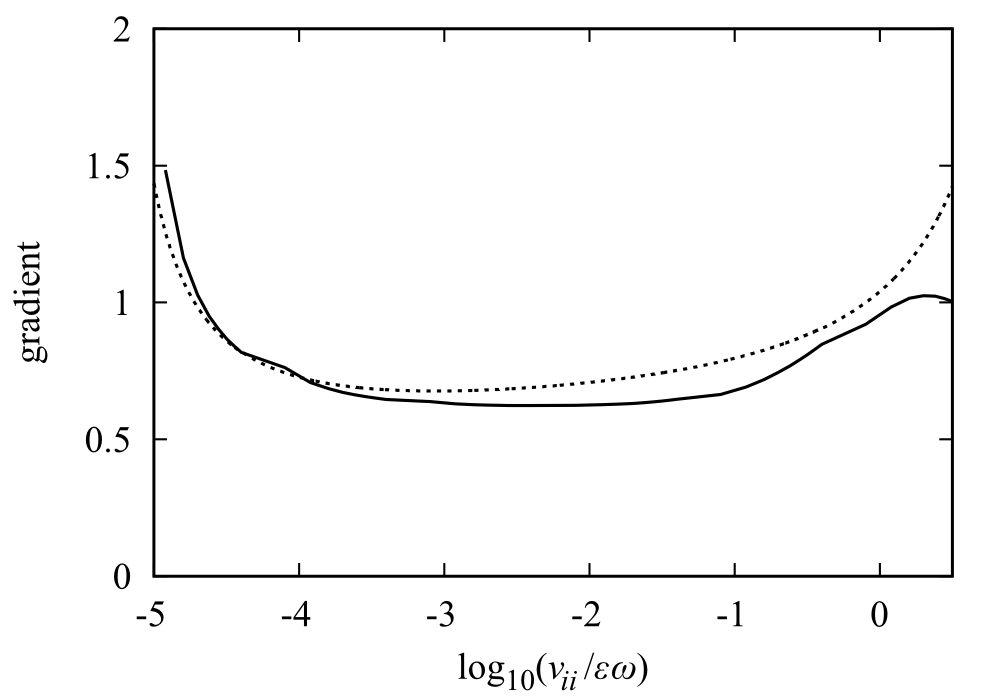}
	\caption{\label{plot2} Plot of the gradient: $d\log(\Deltapol(\nuii)-\Deltapol(\nuii=0)) /d\nuii$ against $\nuii/\epsilon\omega$ for $\epsilon=0.1$, $\eta_{i}=0.5$ and $\omega/\omega_{*i}=2.5$. The dashed line represents the gradient taken from the analytic result in \rref{2009PPCF51-105010}, while the solid line shows the gradient taken from our new numerical result for $g(\nuii,\epsilon,\omega)$. If $g(\nuii,\epsilon,\omega)$ scaled as $\sqrt{\nuii/\epsilon\omega}$, then the gradient is expected to be 1/2. The deviation from 1/2 is due to the weak logarithmic dependence on $\nuii/\epsilon\omega$, predicted analytically \cite{2009PPCF51-105010}.}
\end{figure}

The numerical result for $g(\nuii,\epsilon,\omega)$ as a function of $\nuii/\epsilon\omega$ is shown in \fref{plot1}. As discussed in Section \ref{introduction}, the convention in this paper is that $g>0$ corresponds to a stabilizing contribution to the island evolution. As expected, $g(\nuii,\epsilon,\omega)$ is $O(\epsilon^{3/2})$ smaller in the collisionless limit compared to the collisional limit, and agrees well with the analytic results in both limits. The transition from one limit to another takes place approximately between $\nuii/\epsilon\omega \sim 0.1$ and $\nuii/\epsilon\omega \sim 100$, for $\epsilon =0.1$. Compared to the previous linear MHD results \cite{2000PPCF42-309}, the new results presented here show that $g(\nuii,\epsilon,\omega)$ starts increasing from a somewhat lower value of $\nuii/\epsilon\omega$; in the fitting made to the linear MHD theory in \rref{2000PPCF42-309}, $g$ stays at the low collisionless value until $\nuii/\epsilon\omega \sim1$. This difference is due to the $\sqrt{\nuii/\epsilon\omega}$ dependence of $g(\nuii,\epsilon,\omega)$ arising from the leading order collisional correction in the low collision frequency limit, as predicted by \rref{2009PPCF51-105010}. This $\sqrt{\nuii/\epsilon\omega}$ dependence was omitted in the fitting suggested by \rref{2000PPCF42-309}. However, our results suggest that it is important even when $\nuii/\epsilon\omega \lesssim 1$, enhancing the collisionless neoclassical polarization current by a factor $\sim 2$ at $\nuii/\epsilon\omega \sim 1$, for example (compared to the collisionless value). In \fref{plot2} we show that the leading order collisional correction to $g$ in the collisionless limit is indeed $O(\sqrt{\nuii/\epsilon\omega})$, aside from the weak logarithmic dependence which offsets the gradient from the expected value of 1/2. This is consistent with analytic theory \cite{2009PPCF51-105010}, which predicts such a correction arising from a narrow layer in pitch angle space around the trapped/passing boundary.
%
%

\begin{figure}
	\includegraphics{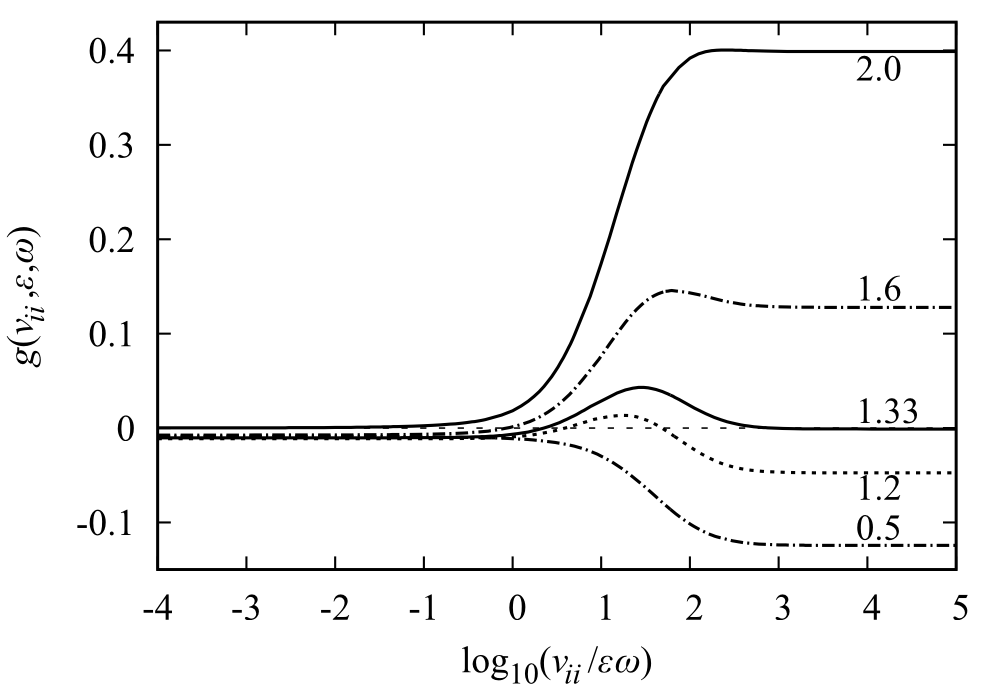}
	\caption{\label{plot3} Plots of $g(\nuii,\epsilon,\omega)$ vs. $\nuii/\epsilon\omega$ for $\epsilon=0.1$ and $\eta_{i} = 1.0$ for different values of $\omega$. The numbers on the right hand side of the graph are the values of $\omega/\omega_{*i}$ for each of the plots.}
\end{figure}

\begin{figure}
	\includegraphics{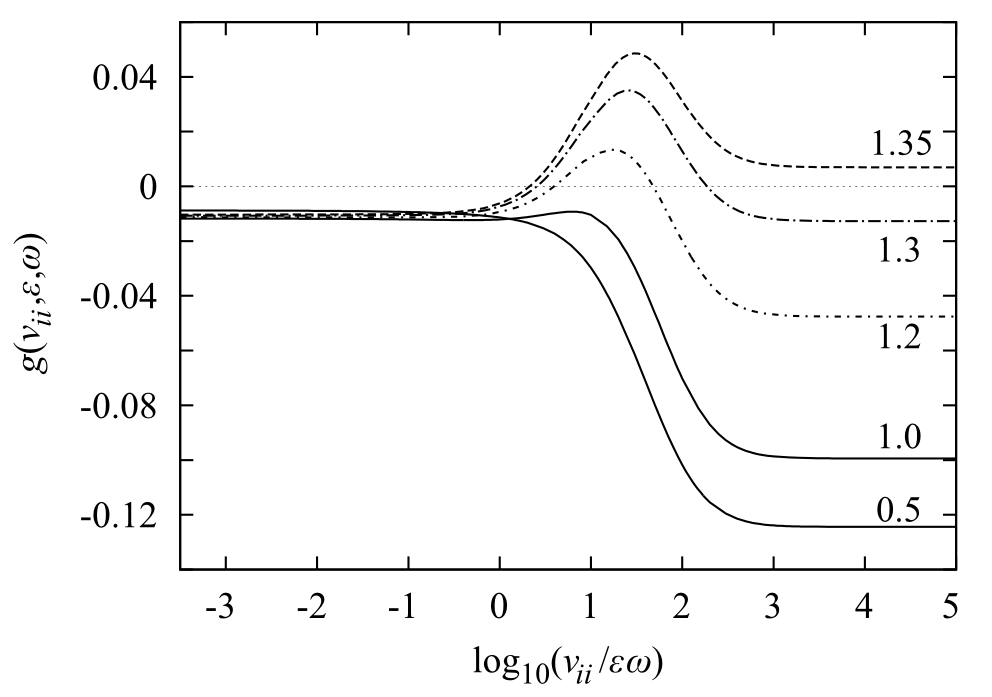}
	\caption{\label{plot4} Plots of $g(\nuii,\epsilon,\omega)$ vs. $\nuii/\epsilon\omega$ for $\epsilon=0.1$ and $\eta_{i} = 1.0$ for smaller values of $\omega$, showing that $g$ is minimum at $\omega/\omega_{*pi}=1/2$ in the collisionless limit, consistent with the analytic form given in \eref{ganalytic}. The numbers on the right hand side of the graph are the values of $\omega/\omega_{*i}$ for each of the plots.}
\end{figure}

\fref{plot1} shows a case where $g(\nuii,\epsilon,\omega)$ is positive (and therefore the external polarization current provides a stabilizing contribution) across all of the collision frequency domain. It turns out, however, that the sign of $g(\nuii,\epsilon,\omega)$ depends on the relative size of $\omega$ with respect to the ion diamagnetic frequency, $\omega_{*i}$. The analytic forms of $g(\nuii,\epsilon,\omega)$ in the collisionless and collisional limits [see \eref{ganalytic}] demonstrate that $g$ is negative if $\omega/\omega_{*i}$ is positive and less than $(1+\eta_{i})$ in the collisionless limit, or if it is less than $1 +(1+k)\eta_{i}$ in the collisional limit. \fref{plot3} shows how $g$ varies with $\omega/\omega_{*i}$ \textit{and} collision frequency. For sufficiently small $\omega$ ($<\omega_{*i}$), $g$ is negative everywhere in the collision frequency domain, and it is positive everywhere for $\omega>\omega_{*i}(1+\eta_{i})$, as expected from \eref{ganalytic}. In \fref{plot4}, with $\epsilon=0.1$ and $\eta_{i}=1.0$, we show that the minimum for $g(\nuii,\epsilon,\omega)$ in the collisionless limit is indeed at $\omega/\omega_{*i}=1$, as expected from the dependence $\omega[\omega-\omega_{*i}(1+\eta_{i})]$, with $\eta_{i} =1.0$. Again, we see that our numerical results are in agreement with the analytic results in the appropriate limits.
%
%

\begin{figure}
	\includegraphics{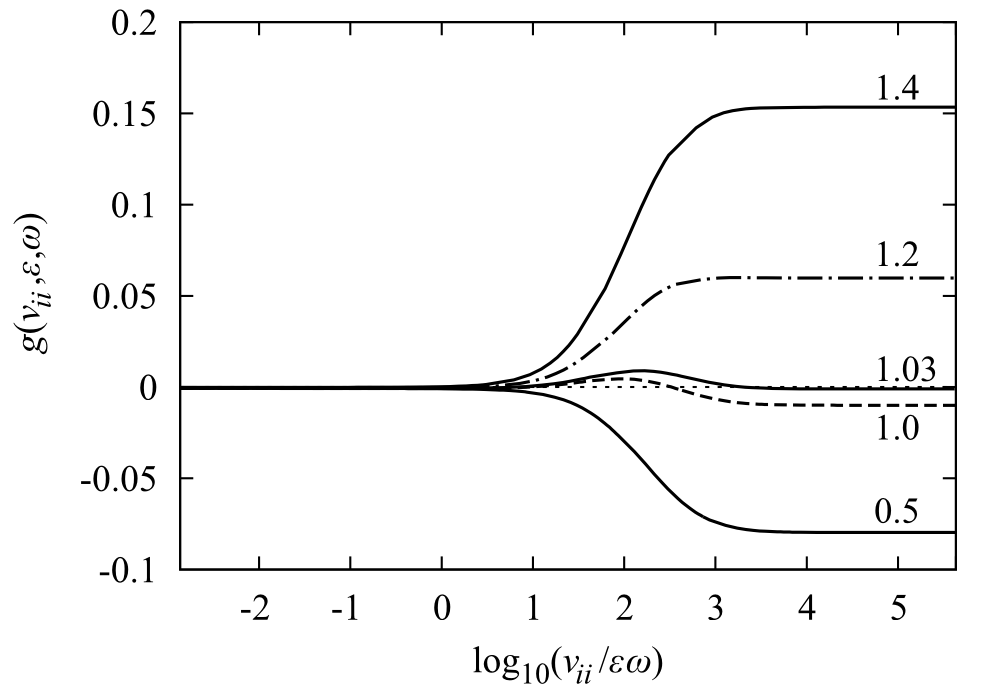}
	\caption{\label{plot5} Plots of $g(\nuii,\epsilon,\omega)$ vs. $\nuii/\epsilon\omega$ for $\epsilon=0.02$ and $\eta_{i} = 0.5$ for different values of $\omega$. The numbers on the right hand side of the graph are the values of $\omega/\omega_{*i}$ for each of the plots.}
\end{figure}

A particularly interesting feature that is evident from \fref{plot3} is that the sign of $g$ changes in the intermediate collision frequency regime as $\nuii/\epsilon\omega$ is increased, when $\omega_{int} < \omega \lesssim \omega_{*i}(1+\eta_{i})$, where for the parameters used in \fref{plot3}, $\omega_{int} \simeq 1.1\omega_{*i}$. Furthermore, there is a range of $\omega$ where $g$ has a maximum at intermediate collision frequencies between the two analytic limits. The implication of this is that whether the neoclassical polarization current can stabilize or amplify the magnetic island depends not only on plasma parameters, including the collision frequency regime, but also on the relative rotation between the island and the plasma. Conversely, whether or not the external polarization current heals or amplifies magnetic islands can depend sensitively on the collision frequency regime. In \fref{plot5} we show that this behavior of $g$ is robust for different values of $\epsilon$ and $\eta_{i}$.
%

Finally, we consider the relationship between the coefficient $k(\epsilon)$ and the critical value of $\omega$ in the collisional limit $\omega_{c}$, for which the sign of $g(\nuii,\epsilon,\omega)$ reverses. According to \eref{ganalytic}, this critical $\omega$ is expected to be: $\omega_{c}= \omega_{*pi} +k(\epsilon)\eta_{i}\omega_{*i}$. Using $k(\epsilon) = -1.17f_{c}(\epsilon)$\cite{helander_sigmar195} and explicitly calculating the passing particle fraction, $f_{c}$, we find $k=-0.67$ and hence $\omega_{c}=1.33\omega_{*i}$, for $\epsilon=0.1$ and $\eta_{i}=1.0$. This is consistent with our numerical result, where we see from Figs. \ref{plot3} and \ref{plot4} that $\omega_{c}=1.33\omega_{*i}$. Lowering $\epsilon$ to 0.02 we find $k=-0.93$ and have $\omega_{c}=1.03$ with $\eta_{i}=0.5$, which is again in agreement with $\omega_{c}$ obtained from the numerical solution of \eref{constraint_passing}, shown in \fref{plot5}. An important point to make is that in both cases, $k(\epsilon)$ is substantially different from the infinite aspect ratio limit: $k(\epsilon=0) =-1.17$, even though $\epsilon$ is very small. The significance of this is that the $f_{c}$-dependence of $k$ should be properly taken into account in any quantitative analysis, even for a very small value of the inverse aspect ratio.

\section{Conclusion}
\label{conclusion}

We have separated the contribution of the neoclassical polarization current to the island evolution into two parts: the \dq{external} polarization current which exists outside the island, and a \dq{layer} polarization current which exists in the complex boundary layer in the vicinity of the island separatrix. We have focused on the external contribution and determined the full collision frequency dependence of its contribution to the island evolution, using nonlinear drift kinetic theory. Our numerical results show that the collisional correction to this external contribution to the neoclassical polarization current is important even at very low collision frequency, $\nuii/\epsilon\omega \sim 0.1$ ($\epsilon=0.1$). Furthermore, we have found a rich structure in the contribution of the polarization current to the NTM evolution in the intermediate collision frequency regime which is not accessible by analytic theory. We find that whether it can provide a stabilizing contribution to the island evolution depends crucially on the size of $\omega/\omega_{*i}$ as well as the collision frequency regime, $\nuii/\epsilon\omega$. Our results suggest that an element of an NTM avoidance or control scheme in future devices, such as ITER, may be through control of the plasma collision frequency as well as the rotation of magnetic islands. Future work will address the layer contribution to the polarization current. This opposes the external contribution that we discussed here, and therefore is important to include in order to make specific quantitative statements concerning the stabilizing influence of total neoclassical ion polarization current on NTMs.

\begin{acknowledgments}
	This work was funded by the Engineering and Physical Sciences Research Council grants 	EP/D065399/1 and EP/H049460/1. The authors would like to thank Jack Connor and Jim Hastie for helpful discussions.
\end{acknowledgments}


\end{document}